\documentclass{llncs}

\usepackage[margin=1.4in]{geometry}
\usepackage{graphicx,times,color}
\usepackage{times}
\usepackage{booktabs}
\usepackage[final]{pdfpages}
\usepackage{subfigure}
\usepackage{amssymb}
\usepackage{url}

\newcommand{\cut}[1]{}

\title{Who clicks there!:\\Anonymizing the photographer in a camera
    saturated society}
\author{ Peter Schaffer\inst{1} \and Djamila Aouada\inst{1}  \and Shishir Nagaraja \inst{2}}

\institute{Interdisciplinary Centre for Security, Reliability and Trust,\\ University of Luxembourg
\and
IIIT Delhi, India}

\begin{document}
\maketitle

\abstract{In recent years, social media has played an increasingly
  important role in reporting world events. The publication of crowd-sourced photographs and videos in near real-time is one of the
  reasons behind the high impact. However, the use of a camera can
  draw the photographer into a situation of conflict. Examples include
  the use of cameras by regulators collecting evidence of Mafia
  operations; citizens collecting evidence of corruption at a public
  service outlet; and political dissidents protesting at public
  rallies. In all these cases, the published images contain fairly
  unambiguous clues about the location of the photographer (scene
  viewpoint information). In the presence of adversary operated
  cameras, it can be easy to identify the photographer by also
  combining leaked information from the photographs themselves. We
  call this the {\bf camera location detection} attack. We propose and
  review defense techniques against such attacks. Defenses such as
  image obfuscation techniques do not protect camera-location
  information; current anonymous publication technologies do not help
  either. However, the use of view synthesis algorithms could be a
  promising step in the direction of providing probabilistic privacy
  guarantees.}

%

\section{Introduction}
Cameras are becoming ubiquitous.  In the near future almost everyone
will carry a personal high quality camera included in a mobile
computing device. Furthermore, surveillance cameras are also
increasing in spaces frequented by the general public. In the last few
years, the quality of such cameras has significantly increased with an
increasing number of deployments on the streets. In the following
paragraphs we highlight novel threats to privacy when adversary
operated cameras point at other camera users.

In a world where cameras are ubiquitous the privacy of individuals is
constantly at stake. The release of pictorial information both careful
and careless has historically been considered a potential privacy risk
based on the leakage of information contained in the picture itself
(private information contained in the captured scene). In this context
the use of ad-hoc obfuscation methods such as blurring, pixelation and
inpainting algorithms~\cite{alg:blurring} have been
proposed. Subsequent work offered privacy metrics and proposed better
solutions on images~\cite{deident-image} and
videos~\cite{deident-vid}.

Little attention has been paid to a possibly more important threat in
the context of camera ubiquity: the privacy of the photographer taking
an image (privacy of location from which the shots have been
taken). Indeed, the average camera person expects to remain anonymous
when publishing an image via anonymous channels on an anonymous
publishing system, however, this is not the case as images leak potentially private information that can lead to the de-anonymization of the photographer.

As a motivating example consider a secret meeting where the Mafia is
announcing new rates for getting access to various public
services. The Mafia's security keep a tight eye on the gathering to
prevent any overt photography. They setup CCTVs to cover the venue but
do not actively monitor the footage. A few journalists in the guise
of businessmen are carrying covert cameras to record the event and
promptly publish their images. When the embarrassing pictures turn up
on the Internet, the Mafia uses the published images to locate the set
of possible camera locations and then leverage CCTV input to identify
the photographer (or at least to significantly minimize the anonymity
set).

This is a real threat faced by increasing numbers of members of the
general public. During the Burmese emergency of
2007~\cite{burma-shutdown}, protestors taking photographs at explosive
anti-government rallies were identified and hunted down by crews of
government secret police collecting footage of protestors in
attendance using mobile video cameras in a bid to censor 'sensitive'
footage. Similar video intelligence teams have since been funded by
several state and non-state actors. We foresee the usage of hostile cameras along with
camera location deanonymization as a serious threat to parties
engaging in activities involving taking pictures in hostile
environments. 

Our goal is to develop methods and techniques to address leakage of
camera location information. Essentially, we hope to investigate novel
techniques of image processing that seek to maximize the privacy of
the camera user or, to use a term from anonymous communications
parlance, to maximize the \emph{unlinkability} between an image and
the photographer.
\section{Problem Description, Threat Model and Solution Approach}
In this paper, we wish to discuss solutions to the problem of
photographer identification or photographer deanonymization, i.e., given
a photograph we wish to anonymize the photographer. In particular, we
focus on maximizing camera-location anonymity.

\paragraph{Camera location anonymity problem:} Given a set of people
 $P$ within a scene where $k \leq |P|$ people are recording (not
necessarily from the same location) and publishing images to the
Internet. We wish to maximize the \textit{unlinkability} between a
published image ({\bf output}) and the location from which it was
captured ({\bf input}). Anonymity is achieved by minimizing the
correlation between raw input and published images. This is similar to
the function of a mix in anonymous communications, where the
correlation between input and output traffic streams is minimized.

\paragraph{Attack:} Given a published image, an attacker 
can infer on the set of potential locations from which an image
could have been captured by examining the scene viewpoint. This is
because a given camera at a scene will record the image of the scene
from a certain viewpoint. In this way, the recorded image will reveal
some information about the location of the camera.  An adversary can
mount a camera location detection attack by combining information from
a leaked (or published) image and footage from adversary cameras
operating within the scene.

The image may also reveal information specific to the camera itself
such as aberrations in the lens or the CCD which can be used to
specifically identify the camera used in capturing a scene~\cite{Choi,Popescu,Lukas}. Such
`side-channel' information can subsequently be used to identify the
photographer. However, we do not seek to defend against such attacks in this
work.

%

\paragraph{Threat model:} 
Our threat model is that of a global passive adversary who deploys
surveillance probes (video cameras). However, we assume that the
adversary does not have the resources to analyze camera footage in
real-time but rather that all video analysis is post-event.

\paragraph{Approach:}
Our approach towards maximizing camera location is the following.
Instead of publishing an image downloaded from the camera's memory
card, the photographer records multiple images (or frames) of the
scene of interest from different viewpoints. She then generates a new
image from a randomly chosen viewpoint using the input images and
publishes the result. Since the scene viewpoint could potentially point
anywhere within the physical space of interest within limits, the
photographer's anonymity can hopefully be maximized to the entire set
of people present in that space. A view synthesis algorithm takes two
images and an input viewpoint and generates a synthesized image from
that viewpoint using the input images. There is a large body of
literature on this in the area of computer vision. View synthesis
depends on a stereo correspondence algorithm applied on the input
images followed by forward or inverse warping steps and a
hole-handling step~\cite{rogmans:viewsyn}. Each of these steps
corresponds to a sub-algorithm. For a detailed review of view
synthesis the reader is referred to the work of Scharstein and
Szeliski~\cite{scharstein:ijcv2002}.
%
%
%
%
%
%
%
%

%
\section{Privacy Analysis}

Let $\mathbf{I}_L$ and $\mathbf{I}_R$  be two images  of the same scene but taken from two different viewpoints, respectively left and right, and corresponding to the two camera positions $L=(x_L, y_L, z_L)$ and $R=(x_R, y_R, z_R)$. The objective of a view synthesis algorithm is to compute a synthesized image $\mathbf{I}_S$ at a new viewpoint or camera position $S=(x_S, y_S, z_S)$. This new image aims at protecting the anonymity of the photographers.  In what follows, we investigate to which extent this anonymity can be preserved. Concretely, given a synthesized image $\mathbf{I}_S$, how much can be inferred about the photographers positions $L$ and $R$ ?\\
To answer this question, we start by identifying at which level privacy can be leaked in a classical stereo-based view synthesis algorithm. Such an algorithm starts by establishing a stereo correspondence between $\mathbf{I}_L$ and $\mathbf{I}_R$. This means estimating a disparity map $\mathbf{D}$ that will enable relating\footnote{For simplicity, we assume a disparity change on the $x$ axis only.} the pixel coordinates $\mathbf{p}_L^T=(u_L, v_L)$ on $\mathbf{I}_L$ to the pixel coordinates $\mathbf{p}_R^T=(u_R, v_R)$ on $\mathbf{I}_R$, such that,
\begin{equation}
 (u_R, v_R)=(u_L + \mathbf{D}(u_L, v_L) , v_L).
\label{eq1} 
\end{equation}
For a realistic rendering of the synthesized image $\mathbf{I}_S$, it is necessary to find a dense disparity map that relates all pairs of points $(\mathbf{p}_L, \mathbf{p}_R)$. To that end, it is necessary to find the most optimal match between $\mathbf{I}_L$ and $\mathbf{I}_R$. If this correspondence is perfectly achieved, the only possible errors can only be due to geometrical constraints such as occlusions. These errors are contained in $\mathbf{D}$, and often translated on $\mathbf{I}_S$ as \emph{holes}. The location of these holes may strongly infer on the relative geometry of the scene as seen from $L$ and $R$; which implies inferring on $L$ and $R$ if the scene geometry is known or can be discovered.

Once $\mathbf{D}$ is estimated, the image $\mathbf{I}_S$ can be computed by interpolation or extrapolation, depending whether $x_S$ is chosen to be inside or outside the interval $[x_L, x_R]$. Hence, at a given pixel location $(u,v)$, and for a baseline $b=(x_R-x_L)$, $\mathbf{I}_S$  may be written as:
\begin{equation}
\mathbf{I}_S (u,v)= \left(1-\left(\frac{x_S-x_L}{b}\right)\right)\cdot\mathbf{I}_L\left(u + \mathbf{D}(u, v) , v\right) + \left(\frac{x_S-x_L}{b}\right)\cdot\mathbf{I}_R (u,v).
\label{eq2} 
\end{equation}
In what follows, we investigate the nature and the extent of the occlusions that cause privacy leakage for the two distinctive cases of extrapolation and interpolation.

\subsection{Privacy Leakage on Extrapolated Records}
\label{Extrapolation}

The observation this section is based upon is that independently of the object recorded there may be occlusion beside the object on the extrapolated record. The reason for this occlusion is that, in specific setups, neither of the two original cameras can record what is behind the object as the object itself occludes that part of the scene. Now, if one is trying to establish an extrapolated record based on such two original records, then the final record will contain the projection of the occluded scene part. The projection of the occluded scene part is called \emph{hole}. The existence of holes is independent of the effectiveness of the stereo correspondence subroutine of the view synthesis algorithm. Even in case of perfect matching, the occluded scene parts cannot be mitigated from the synthesized record. Of course, view synthesis algorithms incorporate a hole filling step, but this is usually based on some kind of interpolation between the hole surroundings, and as such, filled holes are easily observable and measurable by humans (see, for example, Fig.~1. in~\cite{Fusiello}). In the following, we assume that the observer is given an extrapolated picture containing the holes. The observer (or big brother) is trying to recover the position of the original cameras, thus trying to de-anonymize the photographers using the side-channel information given by the position and size of the holes.

In fact, there are two slightly different setups when holes appear on the synthesised picture. These setups are characterized by the
\begin{itemize}
	\item Position of the left camera $L$, position of the right camera $R$ and the distance between them $b$.
	\item Position $(x_S, y_S, z_S)$ of the synthesized viewpoint $S$, and its distance $s$ from $R$.
	\item Object size $l$ and its position, particularly its distance $Z_o$ from the focal plane and size $\hat{l}$ of its projection on the image plane.
	\item Distance $Z_b$ of the background from the cameras, and the focal plane.
	\item Position of the planes (i.e., focal plane, image plane), particularly the focal length $f$.
	\item Hole size $h$ on the image plane.
\end{itemize}

\begin{figure*}[!ht]
	\centering
	\subfigure[First setup giving $h'$]{
		\includegraphics[width=0.80\textwidth]{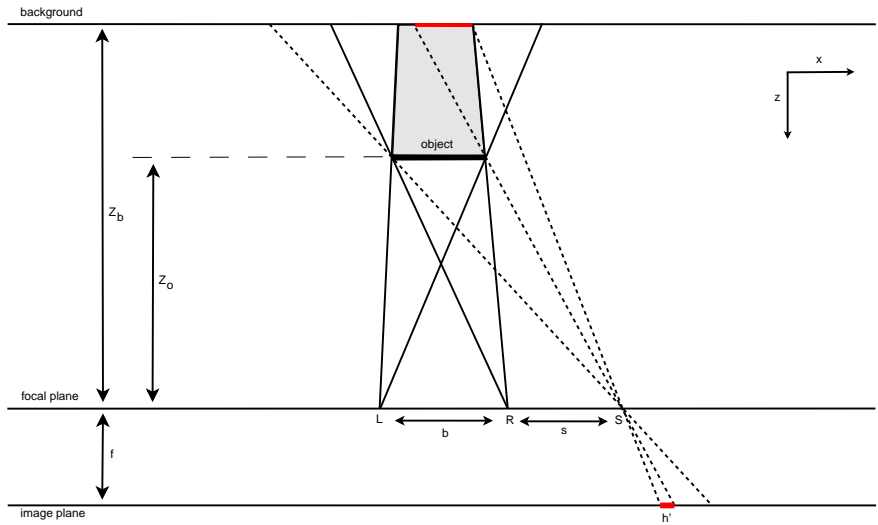}
	  \label{fig:case_1}}
	\subfigure[Second setup giving $h''$]{
		\includegraphics[width=0.80\textwidth]{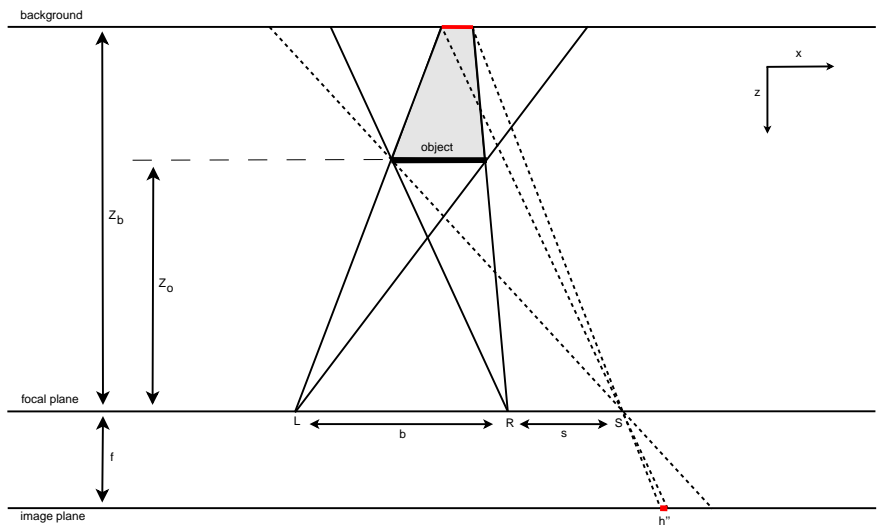}
	\label{fig:case_2}}
	\caption{The two different setups that result in holes on the synthesized picture}
\end{figure*}

Fig.~\ref{fig:case_1} and Fig.~\ref{fig:case_2} depict the two slightly different setups when holes appear. In this simplified scenario the focal plane, the image plane and the background are parallel, and the cameras' fields of view are assumed to be 180$^{\circ}$. The object to record (depicted by the thick black line in the middle) is also parallel to the planes. The grey regions in Fig.~\ref{fig:case_1} and Fig.~\ref{fig:case_2} show the part of the scene that is not seen by either of the cameras $L$ and $R$. In other words, these are the occluded regions. As long as there is no object to record in these regions, no information will be lost because of the occlusion. However, as the background also has an occluded part, a hole will appear as the projection of these occluded parts (which are highlighted in red on the background planes) on the image plane. These holes are denoted by $h'$ and $h''$, respectively, highlighted in red on the image planes (see  Fig.~\ref{fig:case_1} and Fig.~\ref{fig:case_2}).

Note that, in this example we only consider the case when $S$ is to the right from $R$. We do not lose on generality with this assumption, as $S$ being to the left from $L$ would result in the same derivation. Note that there is also a third possible setup, namely when the left ray of camera $L$ and the right ray of camera $R$ are intersecting each other before the background. In this case, however, the background has no occluded part and therefore no hole will appear on the image plane. We will return to this case later on, after having investigated the cases when holes appear.\\
\noindent $h'$ and $h''$ can be calculated with the help of coordinate geometry. By attaching coordinates to the points defining the projection lines of the cameras, $h'$ and $h''$ will take the following forms (note that $h', h''\geq 0$ when holes appear):
\begin{eqnarray}
h' &=& f\bigg( \frac{s}{Z_o}-\frac{s}{Z_b}\bigg) = s\frac{\hat{l}}{l} \bigg( 1-\frac{Z_o}{Z_b} \bigg),\\
h'' &=& f\bigg( \frac{l}{Z_o}-\frac{b}{Z_o}+\frac{b}{Z_b} \bigg) = \hat{l} - b\frac{\hat{l}}{l} \bigg( 1-\frac{Z_o}{Z_b} \bigg).
\end{eqnarray}
Based on $h'$ and $h''$, we can already formulate the general equation for $h$, namely,
\begin{equation}
h = \left\{ \begin{array}{ll}
\min(h',h'') & \textrm{if}\ h'' > 0\\
0 & \textrm{if}\ h'' \leq 0
\end{array} \right.
\label{eqn:h_long}
\end{equation}
By introducing
\begin{equation}
\alpha = \frac{\hat{l}}{l} \bigg( 1-\frac{Z_o}{Z_b} \bigg), 
\end{equation}
$h$ can be rewritten as
\begin{equation}
h = \left\{ \begin{array}{ll}
\min(\alpha s,\hat{l}-\alpha b) & \textrm{if}\ h'' > 0\\
0 & \textrm{if}\ h'' \leq 0
\end{array} \right.
\label{eqn:h_short}
\end{equation}
After this derivation, the question is what effect does the above has on the privacy of the collaborating photographers. In other words, we are interested in the extent to which $s$ and $b$ can be infered on from Eq.~(\ref{eqn:h_short}). The knowledge of $s$ and $b$ together would reveal the photographers directly because, obviously, the virtual coordinates of the synthesized record's focal point are known (from the synthesized image itself) and $s$ and $b$ would give the locations of $L$ and $R$. Naturally, this would mean that the privacy (or more precisely, anonymity) of the photographers is zero. If either $s$ or $b$ is unknown, then the anonymity of the photographers is higher. If both $s$ and $b$ are unknown, then the anonymity is 1. In the following, we will quantify the anonymity of the photographers more precisely based on information leaked by Eq.~(\ref{eqn:h_short}).\\
For the further analysis, we assume that $h$ and $\hat{l}$ are known. The magnification of both $h$ and $\hat{l}$ are measurable on the synthesized picture, therefore the knowledge of their original value depends only on the knowledge of the CCD/CMOS size of the original cameras (as $h$ and $\hat{l}$ are measured on the image plane, which is the CCD/CMOS in this case). There are only a few different CCD/CMOS sizes; the typical CCD/CMOS size for professional cameras is 36$\times$24 mm. We note, however, we do not know whether the value of $h$ is the representation of $h'$ or $h''$. We further assume that $l$ and $f$ can be guessed by the observer. $l$ is the length of the object which, in most of the cases, has well-known dimensions. For example, if a speaker is recorded then $l$ is about 45-55 cm (measured at the shoulders), depending on the gender. The value of $f$ (i.e., the focal length) is guessable by knowing that specific scenes require specific $f$ values. Still considering the example with the speaker, the optimal focal length for capturing him/her would be between 85 and 100 mm, as this is the focal length interval most suitable for (full-length) portraits. Finally, we also assume that $Z_o$ and $Z_b$ are guessable as well. The guessability of the latter two parameters is an implication of the previous assumptions, namely of the guessability of $l$, $\hat{l}$ and $f$. We note, however, that $Z_b$ can only be guessed if the background is not textureless.

Now, from Eq.~(\ref{eqn:h_short}) we know that
\begin{equation}
\textrm{i)}\ h = \alpha s \ \ \ \ \textrm{or}\ \ \ \ \textrm{ii)}\ h = \hat{l}-\alpha b ,
\label{eqn:h_cases}
\end{equation}
but we do not know which of the two cases holds. If the first case prevails then
\begin{eqnarray}
s &=& \frac{h}{\alpha},\\
b &<& \frac{\hat{l}-h}{\alpha},
\end{eqnarray}
which means that the observer knows the position of $R$ precisely and has an upper bound for $b$ (i.e., for the distance between $L$ and $R$). Otherwise, if the second case prevails then
\begin{eqnarray}
s &>& \frac{h}{\alpha},\\
b &=& \frac{\hat{l}-h}{\alpha},
\label{eqn:b_equal}
\end{eqnarray}
meaning that we have a lower bound on the distance of $R$ from $S$ and we know $b$ precisely. 

In the following, we assume that $n$ photographers are positioned with their cameras along a section of the focal plane. We refer to the two ends of this section as $P$ and $Q$, left to right, with coordinates $(x_P,z_P)$ and $(x_Q,z_Q)$, respectively. Note that the worst case from the observer's point of view who is aiming at de-anonymization is when $s$ does not restrict the anonymity set of the photographers, i.e., when the knowledge of $s$ does not exclude any of the suspected photographers. This happens if, in the first case, $x_S-s \geq x_P+b$, and if, in the second case, $x_S-x_Q>s$. In the further analysis we will assume this worst case, i.e., the results at the end will be conservative from the observer's point of view.

By not assuming a single mandatory position for the cameras relative to the photographers body (i.e., the camera is not necessarily located at the centerline of the torso), there are $\big\lceil \frac{b}{l} \big\rceil$ possible pairs of journalists who are suspicious for being the original recorders in the first case. This is because, in the first case, $R$ is known and $L$ is within distance $b$ to $R$. If we consider that the journalists are standing shoulder-to-shoulder with width $l$, the above statement becomes clear. In the second case, there are $\big(n- \big\lceil \frac{b}{l} \big\rceil \big)+\big(n- \big\lfloor \frac{b}{l} \big\rfloor \big)$ possible pairs of suspicious journalists. 

We can now quantify the anonymity of the photographers following~\cite{Diaz} as
\begin{equation}
A = \frac{H(X)}{H_{max}},
\end{equation}
where $H$ stands for entropy. In our case, i.e., when all the pairs within the anonymity set are equally suspicious, the latter equation can be simplified as
\begin{equation}
A = \frac{-\sum_{i=1}^{N_{susp}}\frac{1}{N_{susp}} \log \frac{1}{N_{susp}} }{-\sum_{i=1}^{N}\frac{1}{N} \log \frac{1}{N}} = \frac{\log \frac{1}{N_{susp}}}{\log \frac{1}{N}} = \log_{N} N_{susp},
\label{eqn:anon_def}
\end{equation}
where $N=\frac{n(n-1)}{2}$ is the number of possible pairs of photographers.

When calculating anonymity, the two cases in Eq.~(\ref{eqn:h_cases}) have to be considered together but without their intersection calculated twice. Therefore, having $n$ photographers results in
\begin{equation}
N_{susp}^{\scriptscriptstyle{(h>0)}} = \Big\lceil\frac{b}{l}\Big\rceil + \Big(n-\Big\lceil\frac{b}{l}\Big\rceil\Big)+\Big(n-\Big\lfloor\frac{b}{l}\Big\rfloor\Big) - 1 = 2n - \Big\lceil\frac{b}{l}\Big\rceil,
\label{eqn:anon_hole}
\end{equation}
suspicious pairs (in other words, the size of the anonymity set is $N_{susp}^{\scriptscriptstyle{(h>0)}}$) in case when $h>0$, where $b$ can be calculated as
\begin{equation}
b = \frac{\hat{l}-h}{ \frac{\hat{l}}{l}\big(1-\frac{Z_o}{Z_b}\big) }.
\label{eqn:b}
\end{equation}
As the components of these equations are known or guessable, the anonymity of the photographers can be evaluated using Eq.~(\ref{eqn:anon_hole}) in case $h>0$.

In fact, even the non-existence of holes reveals private information. In order to mitigate holes on the synthesized picture, $L$ and $R$ must reside on one of the specific constellations resulting in no occlusion; and the number of such constellations is geometrically limited. Such constellations are characterized by the fact that the left ray of camera $L$ and the right ray of camera $R$ have to intersect each other before the background plane in order to avoid meaningful occlusions. In such cases $h''$ can be negative.

In the latter case, i.e., when $h=0$, then either $h' \leq 0$ or $h'' \leq 0$. Since $h'$ is always positive, the former implies that
\begin{equation}
b \geq \frac{\hat{l}}{\alpha},
\label{eqn:b_more}
\end{equation}
which is the straightforward opposite of the previous cases when $h''$ was positive. Relying on the previous calculations, the number of suspicious pairs in this case is as follows:
\begin{equation}
N_{susp}^{\scriptscriptstyle{(h=0)}} = \sum_{k=0}^{n-\big\lceil\frac{b}{l}\big\rceil} \bigg(n- \bigg\lfloor \frac{b}{l} \bigg\rfloor -k \bigg).
\label{eqn:Ah0_long}
\end{equation}
The summation in Eq.~(\ref{eqn:Ah0_long}) is justified by the observation that the $h=0$ case is similar to the above second case when $b$ was known and $s$ did not convey information in the worst case to the observer. Eq.~(\ref{eqn:b_more}) does not reveal any hint on $s$, and it can be rewritten as a sum of specific $b$ values, expressed in a discrete way with the help of $k$ when formulating anonymity. Finally, by simplifying Eq.~(\ref{eqn:Ah0_long}), one gets the following expression for the number of suspicious pairs when $h=0$:
\begin{equation}
N_{susp}^{\scriptscriptstyle{(h=0)}} = \Bigg( \frac{n-\big\lceil\frac{b}{l}\big\rceil+2}{2} \Bigg)\Bigg( n-\bigg\lceil\frac{b}{l}\bigg\rceil+1 \Bigg) = N - N_{susp}^{\scriptscriptstyle{(h>0)}},
\label{eqn:anon_no_hole}
\end{equation}
where $b$ can be calculated with the help of Eq.~(\ref{eqn:b}) considering that $h=0$. Anonymity can be further evaluated using Eq.~(\ref{eqn:anon_def}).

To give an example, let us assume a press conference scenario with a speaker and $n$ journalists. The speaker is facing the journalists who are aligned beside each other, two of them stealthily recording the speech. Later these two will collaboratively establish a synthesized, extrapolated record with or without holes beside the speaker depending on the geometrical setup. In this scenario, one can calculate the anonymity of the photographers using Eqs.~(\ref{eqn:anon_def}), (\ref{eqn:anon_hole}) and~(\ref{eqn:anon_no_hole}). Some typical values of parameters required for the evaluation could be as follows:
\begin{itemize}
	\item $n = 20$ (number of journalists)
	\item $l=0.5$ m (length of object, in this case shoulder width)
	\item $\hat{l}=5$ mm (length of projection of the object on the CCD/CMOS)
	\item $Z_o=5$ m (distance between the focal plane (where the journalists are standing) and the speaker)
	\item $Z_b=7$ m (distance between the focal plane and the background)
	\item $h = 1$ mm (size of the hole beside the speaker measured on the CCD/CMOS if there is any, otherwise $h=0$)
\end{itemize}
With these values, the anonymity of the photographers is
\begin{equation}
A = \left\{ \begin{array}{ll}
0.688 & \textrm{if}\ h > 0\\
0.958 & \textrm{if}\ h = 0
\end{array} \right.
\end{equation}
This tells us that the existence of holes reveals a large amount of position information (the anonymity of the photographers is reduced from 1 to 0.688), but even the non-existence of holes is transpiring some private information. According to~\cite{Diaz}, anonymity is considered to be preserved when, roughly, $A\geq 0.8$. Based on this, we can conclude that in case holes appear on the synthesized picture, the level of anonymity of the photographers is not adequate.

\subsection{Privacy Leakage on Interpolated Records}
\label{Interpolation}

\begin{figure*}[!ht]
	\centering

		\includegraphics[width=0.8\textwidth]{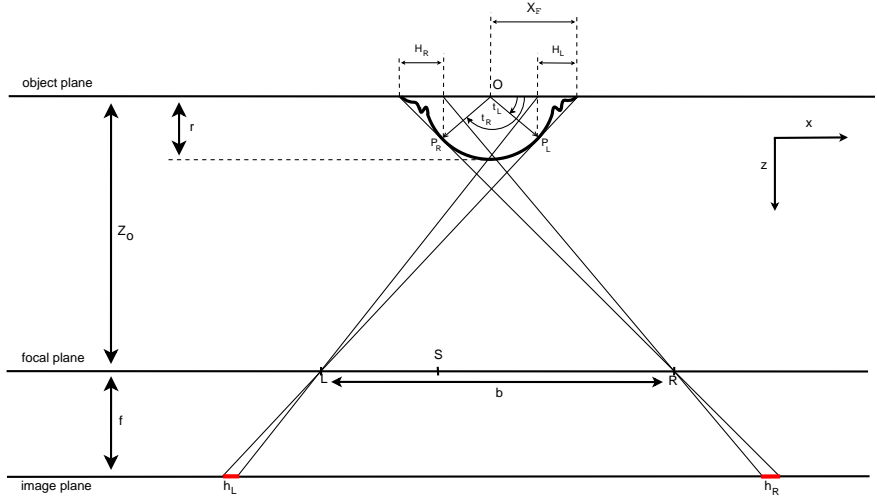}
	  \label{fig:case_3}
	\caption{View synthesis setup in the case of interpolation }
\end{figure*}

In the same way as in the case of extrapolated view synthesis, we assume that the observer is given an interpolated picture $\mathbf{I}_S$. The difference this time is that $x_S\in[x_L, x_R]$. The observer tries to recover the position of the original cameras $L$ and $R$ from the information contained in the holes on $\mathbf{I}_S$. 
We consider the simplified setup depicted in Fig.~\ref{fig:case_3}, with the same assumptions on the cameras as in Section~\ref{Extrapolation}. We model the object to record with the profile depicted by the thick black line, and defined by the function $z=\mathcal{F}(x)$, where the $x$ axis is chosen for simplicity to overlap with the background, and the origin $o=(0,0)$ of the $x$ and $z$ axes approximately coincides  with the center of the support of $\mathcal{F}$, i.e., $\mathcal{F}: x\in[-X_{\mathcal{F}}, X_{\mathcal{F}} ]\longrightarrow\mathbb{R}$. 

The occlusions causing errors on the disparity map $\mathbf{D}$, or in the worst case causing holes, are shown on Fig.~\ref{fig:case_3} as $H_L$ and $H_R$. $H_L$ corresponds to the occluded region on the object relative to the left camera, while $H_R$ is the one relative to the right camera. To find $H_L$, for instance\footnote{The same approach applies to $H_R$.}, we need to find the internal tangent line on $\mathcal{F}$ that intersects with $L$. For simplicity, we assume that this line intersects $\mathcal{F}$  at a single point $P_{L}=(x_{P_{L}}, z_{P_{L}})$. Thus, we may define the line $(LP_L)$ as:
\begin{equation}
(LP_L): z =\mathcal{F'}(x_{P_{L}}) \cdot \left(x-x_{P_{L}}\right) + \mathcal{F}(x_{P_{L}}) ~~\mbox{and}~~
\mathcal{F} \cap (LP_L) = \{P_L\}.
\label{eq3} 
\end{equation}
We use polar coordinates $(t_{L}, r_{L} )$ to define the point $P_L$ such that:
\begin{equation}
x_{P_{L}} = r_{L}\cdot\cos(t_{L})~~\mbox{and}~~z_{P_L} = r_{L}\cdot\sin(t_{L}).
\label{eq4} 
\end{equation}
We approximate the expression of  $\mathcal{F}$ around $P_L$ by a circle centered at the origin $o$, and with a known radius $r_L = r$. Finding $H_L$  becomes equivalent to finding $t_L$. Given $Z_o$, the distance of the object plane from the cameras, we replace the coordinates of $L=(x_L, Z_{o})$ in the equation of $(LP_L)$, and solve for $\sin(t_L)$. We find:
\begin{equation}
\sin(t_L) = \frac{2 r^2 Z_{o} \pm x_L \sqrt{4 Z_{o}^2 + x_L^2 - r^4}}{4 Z_{o}^2 + x_{L}^2},
\label{eq5}
\end{equation}
Therefore:
\begin{equation}
H_L = X_{\mathcal{F}} - r|\cos{t_L}|.
\label{eq6}
\end{equation}
Similarly for the right occlusions, we define $t_R$ using (\ref{eq5}), with $x_L$ replaced by $x_R$. We find:
\begin{equation}
H_R = X_{\mathcal{F}} - r|\cos{t_R}|.
\label{eq7}
\end{equation}
The resulting $H_R$ and $H_L$ correspond to the real sizes of the occlusion regions on the object. The corresponding hole sizes on the image planes are:
\begin{equation}
h_L = \frac{f}{Z_{o}}\cdot\left(X_{\mathcal{F}} - r\cos{t_L}\right) ~~\mbox{and}~~ h_R = \frac{f}{Z_{o}}\cdot\left(X_{\mathcal{F}} - r\cos{t_R}\right).
\label{eq8}
\end{equation}
To evaluate the privacy leakage, one needs to relate the size of the holes on $\mathbf{I}_S$ to $x_L$ and $x_R$. We assume that these holes/errors, that originate from errors on $\mathbf{D}$, do no get further altered by (\ref{eq2}). Based on the position of these holes on $\mathbf{I}_S$, an observer can guess three distinct cases:
\begin{enumerate}
\item \underline{$ (x_R \times x_L) < 0$:} The two cameras are on different sides of the object. The total hole size is consequently $h = (h_L + h_R)$. In this case, $x_L$ and $x_R$ are defined precisely from $t_L$ and $t_R$, respectively. Indeed, by plugging $(x_L, Z_{o})$ and $(x_R, Z_{o})$ in the line equations of $(LP_L)$ and $(RP_R)$, we find; $ x_L = 2\cdot r_L \sin(t_L) - Z_{o}\tan(t_L)$, and  $ x_R = 2\cdot  r_R \sin(t_R) - Z_{o}\tan(t_R)$. This means that positions of the photographers are known, i.e., the number of suspicious pairs of positions $(L, R)$ is $N_{susp}=1$, and consequently from Eq.~(\ref{eqn:anon_def}), anonymity $A = 0$.

\item \underline{$ (x_R \times x_L) > 0$ and $0< x_L < x_R$ :} The two cameras are on the right side of the object. In this case $h=\max(h_L, h_R) = h_R$. This means that only the right camera position $R$ can be known precisely, and for the considered profile $\mathcal{F}$, we have  
\begin{eqnarray}
 N_{susp} &=& \Big\lceil\frac{x_R}{l}\Big\rceil ,\nonumber \\
   &=&  \Big\lceil\frac{1}{l}\big(2\cdot  r_R \sin(t_R) - Z_{o}\tan(t_R)\big)\Big\rceil .
\label{eq9}
\end{eqnarray}

\item \underline{$ (x_R \times x_L) > 0$ and $ x_L < x_R < 0$ :} The two cameras are on the left side of the object. In this case $h=\max(h_L, h_R) = h_L$. This time only the left camera position $L$ can be found precisely, and similarly to Eq.~(\ref{eq9}), for the considered profile $\mathcal{F}$, we find:
\begin{eqnarray}
 N_{susp} &=& \Big\lceil\frac{x_L}{l}\Big\rceil ,\nonumber \\
   &=&  \Big\lceil\frac{1}{l}\big(2 \cdot r_L \sin(t_L) - Z_{o}\tan(t_L)\big)\Big\rceil .
\label{eq10}
\end{eqnarray}

\end{enumerate}

\subsection{Conclusion}
As social media coverage increases and exceeds the coverage of
traditional media, privacy is an increasingly important problem. In
this paper, we have highlighted the threat of camera location detection attacks
mounted by an adversary that combines location clues from published
photographs from
adversary operated cameras in the vicinity of the photographer.  In a
world that is increasingly getting saturated with cameras, this is an
important privacy problem.

Preliminary investigations on analyzing current view synthesis
algorithms indicate decent anonymity gains for modest
computational effort. This indicates a fruitful line of
enquiry in developing defense techniques against camera location
detection attacks and, in turn, defend against the larger class of
photographer de-anonymization attacks.



%

\bibliographystyle{splncs03}
\bibliography{cv}

\begin{thebibliography}{10}
\providecommand{\url}[1]{\texttt{#1}}
\providecommand{\urlprefix}{URL }

\bibitem{deident-vid}
Agrawal, P., Narayanan, P.: Person de-identification in videos. In: Zha, H.,
  Taniguchi, R.i., Maybank, S. (eds.) Computer Vision - ACCV 2009, Lecture
  Notes in Computer Science, vol. 5996, pp. 266--276. Springer Berlin /
  Heidelberg (2010)

\bibitem{Choi}
Choi, K.S., Lam, E.Y., Wong, K.Y.: Automatic source camera identification using
  the intrinsic lens radial distortion. Optics Express  14,  11551--11565
  (2006)

\bibitem{Diaz}
Diaz, C., Seys, S., Claessens, J., Preneel, B.: Towards measuring anonymity.
  In: {Proc. of 2nd Workshop on Privacy Enhancing Technologies (PET)}. San
  Francisco, CA, USA (April 2002)

\bibitem{Fusiello}
Fusiello, A., Colombari, A.: View synthesis along a curve from two uncalibrated
  views. In: {Proc. of the 2nd Workshop On Immersive Communication And
  Broadcast Systems}. Germany (October 2005)

\bibitem{deident-image}
Gross, R., Airoldi, E., Malin, B., Sweeney, L.: Integrating utility into face
  de-identification. In: Danezis, G., Martin, D. (eds.) Privacy Enhancing
  Technologies, Lecture Notes in Computer Science, vol. 3856, pp. 227--242.
  Springer Berlin / Heidelberg (2006),
  \url{http://dx.doi.org/10.1007/11767831_15}

\bibitem{Lukas}
Lukás, J., Fridrich, J., Goljan, M.: Detecting digital image forgeries using
  sensor pattern noise. SPIE Electronic Imaging  (2006)

\bibitem{alg:blurring}
Neustaedter, C., Greenberg, S.: The design of a context-aware home media space
  for balancing privacy and awareness. In: Dey, A.K., Schmidt, A., McCarthy,
  J.F. (eds.) Ubicomp. Lecture Notes in Computer Science, vol. 2864, pp.
  297--314. Springer (2003)

\bibitem{Popescu}
Popescu, A.C., Farid, H.: Exposing digital forgeries in color filter array
  interpolated images. IEEE Transactions on Signal Processing  53,  3948--3959
  (2005)

\bibitem{rogmans:viewsyn}
Rogmans, S., Lu, J., Bekaert, P., Lafruit, G.: Real-time stereo-based view
  synthesis algorithms: A unified framework and evaluation on commodity gpus.
  Image Commun.  24,  49--64 (January 2009),
  \url{http://portal.acm.org/citation.cfm?id=1497635.1497866}

\bibitem{scharstein:ijcv2002}
Scharstein, D., Szeliski, R.: A taxonomy and evaluation of dense two-frame
  stereo correspondence algorithms. Int. J. Comput. Vision  47,  7--42 (April
  2002)

\bibitem{burma-shutdown}
Wang, S., Nagaraja, S.: Pulling the plug: a technical review of the internet
  shutdown in burma. Tech. rep., Harvard University and Cambridge University
  (2007)

\end{thebibliography}

\end{document}